\documentclass[letterpaper,twocolumn,prl,aps,superscriptaddress,amsmath,amssymb,floatfix]{revtex4-1}
\usepackage{ulem}
\usepackage{mathptmx}
\usepackage[latin9]{inputenc}
\usepackage{color}
\usepackage{verbatim}
\usepackage{float}
\usepackage{amsmath}
\usepackage{amssymb}
\usepackage{graphicx}
\usepackage{esint}
\usepackage[unicode=true,
bookmarks=true,bookmarksnumbered=false,bookmarksopen=false,
breaklinks=false,pdfborder={0 0 1},backref=false,colorlinks=true]
{hyperref}
\hypersetup{
	linkcolor=magenta,urlcolor=blue,citecolor=blue,pdfstartview={FitH},hyperfootnotes=false}

\makeatletter

\usepackage{textcomp}
\usepackage{epstopdf}

\pdfpageheight\paperheight
\pdfpagewidth\paperwidth

\@ifundefined{textcolor}{}{%
	\definecolor{BLACK}{gray}{0}
	\definecolor{WHITE}{gray}{1}
	\definecolor{RED}{rgb}{1,0,0}
	\definecolor{GREEN}{rgb}{0,1,0}
	\definecolor{BLUE}{rgb}{0,0,1}
	\definecolor{CYAN}{cmyk}{1,0,0,0}
	\definecolor{MAGENTA}{cmyk}{0,1,0,0}
	\definecolor{YELLOW}{cmyk}{0,0,1,0}
}

\usepackage{xcolor}
\usepackage{soul}
\newcommand{\ket}[1]{\ensuremath{\left|#1\right\rangle}}

\definecolor{blue}{rgb}{0,0,1}
\definecolor{red}{rgb}{0,0,0}
\definecolor{green}{rgb}{0,1,0}
\newcommand{\red}[1]{\textcolor{red}{ #1}}

\usepackage{soul}

\makeatother

\begin{document}
\title{High-performance quantum interconnect between bosonic modules beyond transmission loss constraints}
    \author{Hongwei Huang}
    \thanks{These two authors contributed equally to this work.}
    \author{Jie Zhou}
    \thanks{These two authors contributed equally to this work.}
    \affiliation{Center for Quantum Information, Institute for Interdisciplinary Information Sciences, Tsinghua University, Beijing 100084, China}
	\author{Weizhou~Cai}
    \affiliation{Laboratory of Quantum Information, University of Science and Technology of China, Hefei 230026, China}
	\author{Weiting Wang}
	\affiliation{Center for Quantum Information, Institute for Interdisciplinary Information Sciences, Tsinghua University, Beijing 100084, China}
	\author{Yilong Zhou}
	\affiliation{Center for Quantum Information, Institute for Interdisciplinary Information Sciences, Tsinghua University, Beijing 100084, China}
    \author{Yunlai Zhu}
	\affiliation{Center for Quantum Information, Institute for Interdisciplinary Information Sciences, Tsinghua University, Beijing 100084, China}
    \author{Ziyue Hua}
	\affiliation{Center for Quantum Information, Institute for Interdisciplinary Information Sciences, Tsinghua University, Beijing 100084, China}
	\author{Yifang Xu}
	\affiliation{Center for Quantum Information, Institute for Interdisciplinary Information Sciences, Tsinghua University, Beijing 100084, China}
	\author{Lida Sun}
	\affiliation{Center for Quantum Information, Institute for Interdisciplinary Information Sciences, Tsinghua University, Beijing 100084, China}
 	
	  \author{Juan Song}
	\affiliation{Beijing Academy of Quantum Information Sciences, Beijing, China}
    \author{Tang Su}
	\affiliation{Beijing Academy of Quantum Information Sciences, Beijing, China}
	\affiliation{Hefei National Laboratory, Hefei 230088, China}

    \author{Ming Li}
    \affiliation{Laboratory of Quantum Information, University of Science and Technology of China, Hefei 230026, China}
    \affiliation{Hefei National Laboratory, Hefei 230088, China}

	\author{Haifeng Yu}
    \email{hfyu@baqis.ac.cn}
	\affiliation{Beijing Academy of Quantum Information Sciences, Beijing, China}
	\affiliation{Hefei National Laboratory, Hefei 230088, China}
	\author{Chang-Ling Zou}
	\email{clzou321@ustc.edu.cn}
    \affiliation{Laboratory of Quantum Information, University of Science and Technology of China, Hefei 230026, China}
     \affiliation{Anhui Province Key Laboratory of Quantum Network, University of Science and Technology of China, Hefei 230026, China}
	\affiliation{Hefei National Laboratory, Hefei 230088, China}
	
	\author{Luyan Sun}
	\email{luyansun@tsinghua.edu.cn}
	\affiliation{Center for Quantum Information, Institute for Interdisciplinary Information Sciences, Tsinghua University, Beijing 100084, China}
	\affiliation{Hefei National Laboratory, Hefei 230088, China}
	
\begin{abstract}
Distributed quantum computing architectures require high-performance quantum interconnects between quantum information processing units, while previous implementations have been fundamentally limited by transmission line losses. Here, we demonstrate a low-loss interconnect between two superconducting modules using an aluminum coaxial cable, achieving a bus mode quality factor of $1.7 \times 10^6$. By employing SNAIL as couplers, we realize inter-modular state transfer in 0.8~$\mu$s via a three-wave mixing process. The state transfer fidelity reaches 98.2\% for quantum states encoded in the first two energy levels, achieving a Bell state fidelity of 92.5\%. Furthermore, we show the capability to transfer high-dimensional states by successfully transmitting binomially encoded logical states. Systematic characterization reveals that performance constraints have shifted from transmission line losses (contributing merely 0.2\% infidelity) to module-channel interface effects and local Kerr nonlinearities. Our work advances the realization of quantum interconnects approaching fundamental capacity limits, paving the way for scalable distributed quantum computing and efficient quantum communications.
	\end{abstract}
	
	\maketitle
	
\textit{Introduction.-} Superconducting quantum circuits have emerged as one of the most promising platforms for realizing universal quantum computation~\cite{Kjaergaard2020,Devoret2013}, demonstrating exceptional coherence times~\cite{baqis_t1,Ganjam2024}, precise control capabilities~\cite{baqis_gate,cz_gate}, and scalability~\cite{PhysRevLett.134.090601,Kim2023,Acharya2025}. This platform has achieved major milestones, including quantum advantage over classical computers~\cite{Arute2019,PhysRevLett.134.090601,Kim2023} and quantum error correction (QEC) surpassing the break-even point~\cite{Ofek2016,Sivak2023,Ni2023,Acharya2025}. Despite these remarkable advances, the inherent limitations in qubit fabrication yield and density in single chips necessitate a distributed architecture to further scale up the quantum system, where separate quantum modules are linked through high-performance quantum interconnects~\cite{Gottesman1999, Kimble2008,jiangliang-module,module_theo,module_photon,module_prx}. Microwave photons, which serve as flying qubits, have been successfully employed to realize quantum state transfer and entanglement between distant superconducting qubits through microwave channels~\cite{flyingphoton_eth,flyingphotons_devoret,flyingphoton_oliver,2refregiators}. Significant progress has been made in establishing quantum connections between qubits residing on the same chip through on-chip microwave waveguides~\cite{onchip_zyp,onchip_cleland}, as well as qubits across separated chips through off-chip superconducting microwave cables or waveguides~\cite{1mstationwave_cleland,cable_zyp,Niu2023,caoxi,song2024,qiu}.


For practical quantum computation tasks consuming high-quality entangled qubit pairs, it is demanded to build quantum interconnects with high entanglement generation rate. This capability is fundamentally bounded by the quantum capacity of the microwave channel~\cite{Lloyd1997,Gyongyosi2018,Noh2019}, \red{the highest asymptotic rate (qubits per channel use) at which quantum information can be transmitted reliably using optimal QEC codes}. Apparently, the approaches employing single photons cannot saturate the quantum capacity. For microwave channels where amplitude decay constitutes the dominant error mechanism, two complementary approaches can enhance the interconnect performance: utilizing higher excitation states of microwave modes and reducing photon transmission loss between modules. The first approach leverages the expanded Hilbert space of bosonic modes to implement QEC codes~\cite{Albert2018,Cai2021}, providing intrinsic protection against photon loss and thereby potentially approaching the quantum capacity limit. \red{Besides, the large Hilbert space of a bosonic mode can encode multiple qubits, significantly enhancing the quantum capacity per round of state transfer.} The second approach focuses on minimizing loss in the microwave channel itself, including both insertion losses at module-channel interfaces and propagation losses within the transmission line~\cite{Niu2023,Magnard2020}. \red{By reducing information leakage,  lower channel loss raises the channel's fundamental capacity, which in turn sets a higher target that optimal coding and system design aim to approach.}

\begin{figure*}[t!]
		\centering
		\includegraphics{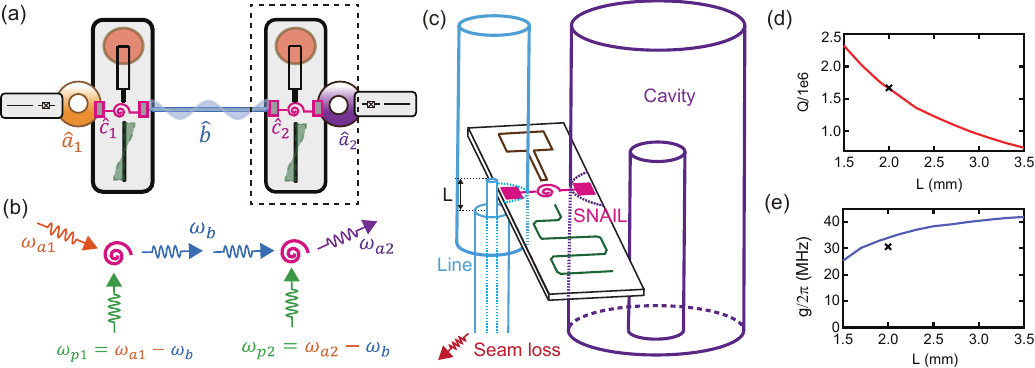}
\caption{(a) Schematic of two superconducting modules connected via a high-quality aluminum coaxial cable as the transmission channel. Each module consists of a 3D cavity, a SNAIL coupler, and a transmon qubit. (b) The three-wave mixing process enabled by the SNAIL. The frequency mismatches between the cavity and the channel mode are compensated by selected driving frequencies. (c) 3D schematic of the module-channel interface, highlighting the peeled length $L$ of the cable for engineering the transmission loss. \red{The energy of the bus mode is dissipated at the clamp-cable interface (not shown) at the bottom of the package. The waveguide above the transmission line is to spatially extend the bus mode and enhance the bus-SNAIL coupling.} (d-e) Trade-off between quality factor $Q$ and coupling strength $g$ versus $L$. The crosses indicate the operating point in our experiment at $L=2$~mm.}
\label{fig:fig1}
\end{figure*}

Bosonic modes in three-dimensional superconducting cavities coupled to transmon qubits offer a hardware-efficient platform for high quantum capacity, enabling unprecedented control over Fock-state superpositions ~\cite{Cai2024,Ofek2016,Ni2023,Hu2019,Ma2020,Sivak2023,Gertler2021,Campagne2020}. This architecture has demonstrated exceptional capabilities for quantum information processing, including the implementation of binomial~\cite{Hu2019,Ni2023} , cat~\cite{Ofek2016}, and GKP error correction codes~\cite{Sivak2023}, along with universal control operations on encoded logical qubits~\cite{cat_universal_rob,Hu2019,Ma2020}. Although pioneering experiments have demonstrated quantum state transfer and entanglement generation using error-corrected bosonic encodings~\cite{Axline2018,error_detected}, their performance remains constrained by transmission line losses, with the quality factor $Q$ restricted to around $10^5$. \red{In contrast, the Q factor of the transmission line is significantly higher in planar qubit systems~\cite{cable_zyp,Niu2023,qiu,caoxi}, primarily attributed to the use of new materials and specific on-chip circuit design.}

	\begin{figure}
		\centering
		\includegraphics{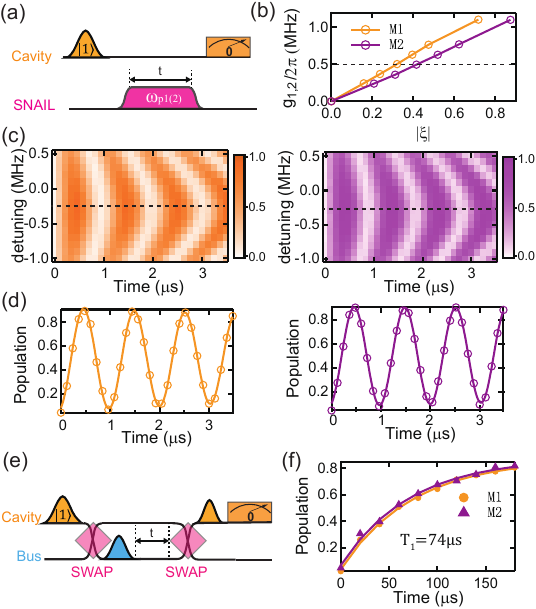}
		\caption{(a) Experimental sequence for characterizing the coherent conversion interaction. (b) Extracted conversion coupling strength under different pump amplitudes ($|\xi|$). (c) Cavity population evolution as a function of pump detuning and interaction time, with dashed lines marking resonant pumping conditions. (d) Line cuts at resonant pumping, with solid lines representing the fitted results for extracting the conversion coupling strength. (e) Pulse sequence for characterizing the bus mode's energy relaxation time. (f) Decay measurements showing single-exponential fits (lines) to the experimental data (dots), yielding $T_1=74\,\mu$s for both modules (module 1: orange; module 2: purple).}
		\label{fig:fig2}
	\end{figure} 

	\begin{figure}
		\centering
		\includegraphics{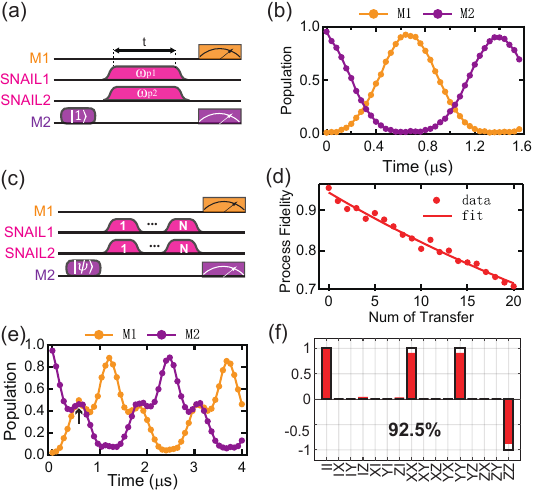}
		\caption{(a) and (b) Pulse sequence and corresponding population dynamics for single-photon transfer between two modules. (c) and (d) Pulse sequence and measured process fidelity for variable numbers of transfer, respectively.
        Solid line: exponential fit, yielding a process fidelity of 98.2\% per state transfer. (e) Time evolution of the populations in the two cavities under detuned pumping $\Delta=\sqrt{8/3}g_\mathrm{BS}$. Arrow at 580\,ns indicates the 50:50 conversion. (f) Joint Pauli measurement results of the two modules (red bar) compared to the ideal value (black frames), demonstrating entanglement fidelity of 92.5\%.}
		\label{fig:fig3}
	\end{figure}

In this work, we use the high-purity aluminum coaxial cables to connect two remote bosonic modules, with carefully optimized coupling geometry. \red{We achieve a bus mode $Q=1.7\times10^6$, which exceeds those achieved in planar qubit systems while maintaining high device interchangeability due to its plug-and-play nature~\cite{caoxi} (see Ref.~\cite{sm} for more details).} Our work reveals a new regime of quantum interconnects where microwave channel losses no longer represent the dominant limitation to quantum state transfer fidelity, as they contribute merely 0.2\% infidelity for transmitting a single photon. In the new regime, the module-channel interface imperfections become the primary constraints: ac-Stark shifts (0.6\% infidelity), SNAIL coupler dephasing (0.3\% infidelity), and cross-Kerr interactions (0.2-0.3\% infidelity). \red{Based on these findings, these parameters can now be further optimized in future designs to achieve state transfer fidelity exceeding 99\%.} We also demonstrate the quantum state transfer with a binomial code and reveal that self-Kerr nonlinearities in superconducting cavities might impose a limitation on quantum interconnect: even kHz-level Kerr coefficients can substantially degrade transfer fidelity when photon numbers increase. Our work establishes a pathway toward high-performance quantum interconnects that approach the fundamental limits of quantum capacity,  opening new frontiers in distributed quantum computation and communication.

\textit{Experimental device.-} Figure~\ref{fig:fig1} shows the schematic of our device, consisting of two modules connected by a 13.5\,cm aluminum coaxial cable that serves as the transmission channel. Each module houses a 3D superconducting cavity~\cite{Reagor2013}, a transmon qubit, and a SNAIL coupler~\cite{Gaoyuan_bs,SNAIL_coupler}. We utilize the fourth mode of the channel ($\omega_b/2\pi=$3.6 GHz) as the bus mode, and coherent conversions between the module and channel is realized through a three-wave mixing process of the SNAIL [Fig.~\ref{fig:fig1}(b)], described by the Hamiltonian
\begin{equation}
   \hat H_\mathrm{BS_{1(2)}}=g_{1(2)}(e^{-i\Delta_{1(2)} t}\hat a^\dagger_{1(2)} \hat b+e^{i\Delta_{1(2)} t}\hat a_{1(2)} \hat b^\dagger).
   \label{eq:BS}
\end{equation}
Here, $\hat a_j$ and $\hat b$ denote the annihilation operators for the mode in module $j$ and the bus mode, respectively. The coupling strength $g_{1(2)}$ is induced by a pump drive on the SNAIL with detuning $\Delta_{1(2)}=\omega_{p1(2)}-\omega_{a1(2)}+\omega_b$, where $\omega_{p1(2)}$ is the pump frequency and $\omega_{a1(2)}$ is the 3D cavity frequency.

The module-channel interface is critical for high performance of the interconnects, with the detailed structure illustrated in Fig.~\ref{fig:fig1}(c). We plug the transmission line from the bottom of the package and peel the outer conductor and dielectric at the end to strengthen the coupling with the SNAIL. As mentioned in previous works, the lifetime of the bus mode is predominantly limited by seam loss at the clamp area~\cite{cable_zyp,error_detected}. According to the numerical simulations in Figs.~\ref{fig:fig1}(d) and \ref{fig:fig1}(e), both the quality factor $Q$ of the bus mode and the SNAIL-bus coupling strength $g$ are controlled by the peeled length of the cable $L$. However, $Q$ and $g$ show a competing relation.It should be noted that SNAIL offers one of the key advantages that a stronger coherent conversion coupling strength $g_{1,2}$ can be achieved for a given $g$ compared to four-wave mixing devices \red{(see Ref.~\cite{sm} for more details)}. We carefully engineer the device and choose an optimized length of $L=2\,\mathrm{mm}$, as marked by crosses in Figs.~\ref{fig:fig1}(d) and \ref{fig:fig1}(e). The bus mode maintains an ultrahigh $Q$ while a large $g$ is experimentally achieved.

To characterize our quantum interconnect, we systematically calibrate the three-wave mixing process in each module separately. Figure~\ref{fig:fig2}(a) shows the experimental calibration procedure: prepare a single-photon Fock state in the cavity with the help of the auxiliary transmon qubit~\cite{Khaneja2005,Heeres2017}, activate the drive on SNAIL for a duration $t$, and then measure the population of the vacuum state (Fock $\ket{0}$) of the cavity by applying a photon-number-selective $\pi$ pulse~\cite{Schuster2007} on the auxiliary transmon. 
Figure~\ref{fig:fig2}(c) shows examples of the population dynamics under varying drive detunings. The coupling strengths $g_{1,2}$ are extracted from the dashed lines with the highest oscillation contrast, as shown in Fig.~\ref{fig:fig2}(d). These extracted $g_{1,2}$ are plotted in Fig.~\ref{fig:fig2}(b) as a function of the dimensionless pump amplitude $\xi = \epsilon/\Delta$, where $\epsilon$ and $\Delta$ represent the pump strength and detuning, respectively. \red{The slight difference between $g_{1}$ and $g_{2}$ is due to fabrication variance of SNAILs.} The experimental results show a linear dependence on the pump amplitude as expected, demonstrating our ability to precisely control the interaction strength up to $1$\,MHz. The complete conversion between the cavity and the bus, i.e., a SWAP operation, is realized by carefully choosing the pump duration. Based on the consequence shown in Fig.~\ref{fig:fig2}(e), the single-photon energy relaxation time of the bus mode is measured to be $T_1=74~\mu$s, yielding a quality factor $Q=1.7\times 10^6$.


\textit{State transfer.-} Following the scheme demonstrated in Ref.~\cite{error_detected}, the interconnection between two modules is tested employing the simplest single-photon encoding. Figure~\ref{fig:fig3}(a) shows the sequence that simultaneously activates the conversions in both modules. Tuning the interaction strengths $g_1=g_2=g_\mathrm{BS}$ and the detuning $\Delta_1=\Delta_2=\Delta$ (see Ref.~\cite{sm} for more details), the system Hamiltonian becomes
\begin{equation}
	\hat H=g_\mathrm{BS} (\hat a^\dagger_{1} \hat b+ \hat a_{1} \hat b^\dagger+\hat a^\dagger_{2} \hat b+ \hat a_{2} \hat b^\dagger)+ \Delta \hat b^\dagger \hat b.
	\label{eq:ST}
\end{equation}
In the following experiments, we choose a moderate $g_\mathrm{BS}/2\pi\approx 0.5\,$MHz to balance the transfer speed and the pump-induced dephasing errors~\cite{dephasing}. Figure~\ref{fig:fig3}(b) displays the single-photon transfer dynamics for $\Delta=0$, demonstrating alternating vacuum population oscillations in the two modules as the transfer duration $t$ is varied. The pump pulses employ two 50~ns smooth edge profiles, which are excluded from the duration $t$. Complete state transfer occurs at $\tau_\mathrm{ST}=\pi/{\sqrt 2 g_\mathrm{BS}}=672$\,ns with an efficiency $\eta= 97.2\%$.


As shown in Fig.~\ref{fig:fig3}(c), quantum process tomography is performed to quantify the fidelity of the inter-modular state transfer. The protocol involves: 1) preparing different initial states in the first two energy levels of the cavity in module 2; 2) repeatedly executing varying number of complete transfer processes; and 3) performing state tomography on the received states. The resulting process fidelity, compared with the identity process matrix, is presented in Fig.~\ref{fig:fig3}(d). An exponential fit reveals the state transfer infidelity of 1.8\%. 
    

Compared to previous works~\cite{Axline2018, error_detected}, our protocol demonstrates significantly enhanced state transfer fidelity, primarily attributed to the extended lifetime of the transmission channel. Our simulations reveal that single-photon loss in the transmission channel contributes merely 0.2\% infidelity, indicating it is no longer the dominant limitation for state transfer fidelity. The remaining infidelity can be categorized as follows: First, cavity decoherence contributes 0.5\% infidelity, which can be further mitigated through sample design optimization. Second, the pure dephasing of SNAIL imposes a limitation on the transmission channel's pure dephasing time, accounting for 0.3\% infidelity. 
Third, the pump during the beamsplitter operation induces thermal excitation in the SNAIL. Since our SNAILs are not operated at the Kerr-free point, residual cross-Kerr interactions between SNAIL, transmission channel, and storage cavity cause frequency shifts during the thermal excitation. This leads to dephasing and contributes 0.2\%-0.3\% infidelity. Finally, the remaining 0.6\% infidelity originates from the ac-Stark shifts induced by the pump. The time-dependent pump amplitude during the ramp-up/down profiles creates varying ac-Stark shifts, violating resonance conditions for the beamsplitter interaction. This effect can potentially be mitigated through dynamic compensation techniques as demonstrated in Ref.~\cite{xyf_dynamic}. 

Using the same Hamiltonian that governs the complete state transfer [Eq.~(\ref{eq:ST})], we tune the detuning $\Delta=\sqrt{8/3}g_\mathrm{BS}$ to create a 50:50 conversion between the two modules. This specific detuning ensures that after the appropriate interaction time, exactly half of the quantum amplitude is transferred between modules, with the bus mode returning precisely to its vacuum state. The population dynamics in Fig.~\ref{fig:fig3}(e) shows that after an evolution time of $t=580$~ns (marked by the arrow), the populations of the two modules equalize. At this time, the two modules should ideally be in the maximally entangled Bell state $\left| \psi \right\rangle =(\left| 01 \right\rangle+\left| 10 \right\rangle)/\sqrt 2 $. To characterize the fidelity of the generated entangled state, we: 1) prepare $\left| 1 \right\rangle$ in module 2, 2) activate the detuned pumps for 50:50 conversion, and 3) perform joint state tomography by decoding the quantum states in the two modules onto the two auxiliary qubits. Figure~\ref{fig:fig3}(f) displays the experimentally reconstructed expectation values of the two-qubit Pauli operators, with ideal theoretical values shown as black frames for comparison. We achieve a Bell state fidelity of 92.5\%, which is primarily limited by the state preparation and measurement (SPAM) error of about 95.2\%~\cite{sm}.

\begin{figure}
	\centering
	\includegraphics{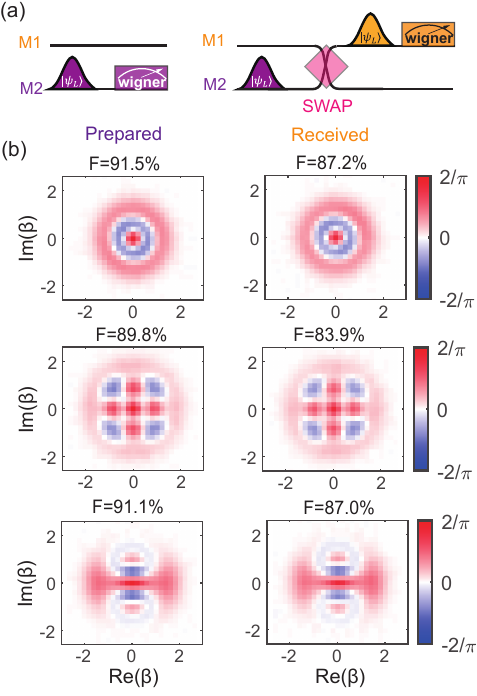}
	\caption{Binomial logical state transfer. (a) Pulse sequence for the binomially encoded state transfer. Left: Wigner tomography is performed immediately after state preparation in module M2 to measure the infidelity due to SPAM errors. Right: After state preparation, the quantum state is transferred to module M1 for Wigner function measurement. By comparing the fidelity before and after the state transfer, we can eliminate the SPAM errors.
    (b) Prepared and received binomial logical states $\left| 0_L \right\rangle$, $\left|1_L \right\rangle$, and $\left|+_L \right\rangle$ that are characterized by Wigner functions. The deterministic phase shift from the state transfer and measurement process is compensated virtually. }
	\label{fig:fig5}
\end{figure}
\begin{figure}
	\centering
	\includegraphics{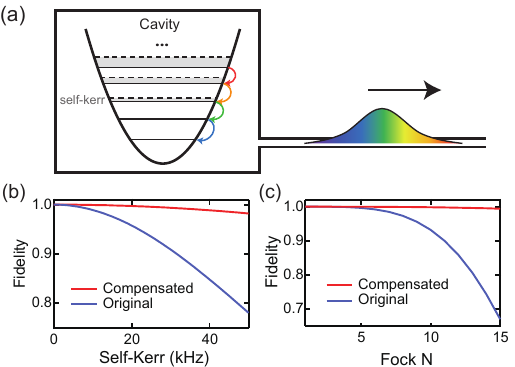}
	\caption{Simulated self-Kerr effect on multi-photon state transfer. (a) Schematic of self-Kerr effect in a cavity. As photon numbers increase, the nonlinear effect becomes more pronounced, inducing photon-number-dependent frequency shifts. These shifts lead to temporally chirped wavepackets after conversion from the module. (b) Average received state fidelity for the binomial code as a function of self-Kerr. The blue curve represents the originally received state fidelity, while the red curve shows the fidelity after applying a corrective phase gate. (c) Received state fidelity for the superposition state $\ket{0}+\ket{N}$ as a function of photon number $N$, using the same self-Kerr strength as in the experimental device~\cite{sm}.}
	\label{fig:fig6}
\end{figure}

	
\textit{Binomial logical state transfer.-} Unlike two-level qubits, the bosonic module offers access to infinite-dimensional Hilbert space to transfer complex and high-dimensional quantum states. To demonstrate the potential of our interconnects for enhanced quantum capacity, we implement the transfer of binomially encoded logical states~\cite{binomial_prx,Hu2019} between modules, focusing on three representative logical states $\left| 0_L \right\rangle=(\left| 0 \right\rangle+\left| 4 \right\rangle)/\sqrt{2}$, $\ket{1_L}=\ket{2}$, and $\left| +_L \right\rangle=(|0_L\rangle+|1_L\rangle)/\sqrt{2}$. The experimental sequence and reconstructed Wigner functions are shown in Fig.~\ref{fig:fig5}. We first prepare the logical states in the cavity of module 2 using a numerically optimized pulse~\cite{Khaneja2005,Heeres2017}. After applying the pumps, we perform Wigner tomography on the received states to get the state fidelity. In order to separate the influence of SPAM errors, we also perform Wigner tomography on the prepared states in module 2. The state fidelities of the prepared and received states are marked in Fig.~\ref{fig:fig5}(b). The average fidelity of the prepared states is $F_1=90.8\%$, while the received states exhibit an average fidelity of $F_2=86.0\%$. Considering that the contributions of SPAM errors from both modules are similar~\cite{sm}, the fidelity of logical state transfer is estimated to be around $F_2/F_1=94.8\%$. Although higher photon numbers (up to n = 4) are involved in transferring these states, our numerical simulation confirms that the contribution of photon loss to infidelity remains remarkably low at approximately 0.6\%. These results further validate that our ultra-high-$Q$ bus mode effectively overcomes the bottleneck of traditional transmission loss. Additionally, our results highlight the contributions of module-channel interface in high-dimensional encodings, including the decoherence in each mode and the pump-induced dephasing (see Ref.~\cite{sm} for more details). 

While self-Kerr effects remain negligible in our current experiments, they emerge as a fundamental limitation for quantum states with higher photon numbers necessary for achieving higher information capacity. As illustrated in Fig.~\ref{fig:fig6}(a), self-Kerr nonlinearities in the cavities become increasingly significant as photon numbers rise, causing photon-number-dependent frequency shift. This shift results in temporally chirped wavepackets after conversion from the module. This phenomenon introduces two key imperfections: First, a frequency mismatch arises between the time-forward chirped pulse emitted by the sending module and the time-reversed chirp expected by the receiving module, degrading the fidelity of the transferred state. Second, photon-loss events correlate with specific frequency components, allowing the environment to extract partial information about the photon-number distribution. This leads to additional dephasing between Fock-state components.

Figure~\ref{fig:fig6}(b) presents numerical simulations of the average transferred state fidelity for the binomial code discussed earlier as a function of the self-Kerr of the cavity. The fidelity decreases rapidly with increasing Kerr value (blue curve), reaching an infidelity around 0.1\% at current experimental values (kHz-level). However, even kHz-scale self-Kerr effects can substantially degrade quantum information encoded in high-photon-number states, as demonstrated in Fig.~\ref{fig:fig6}(c) explicitly: the fidelity of states $\ket{0} + \ket{N}$ at the receiver decreases with increasing $N$. We find that applying a compensated phase gate at the receiver can effectively mitigate these Kerr-induced errors, as shown by the red curves in Figs.~\ref{fig:fig6}(b) and \ref{fig:fig6}(c). Additional mitigation strategies, such as the ``PASS" technique demonstrated in Ref.~\cite{Ma2020} or numerically optimized control pulses may also be helpful and worth further investigations. 


\textit{Conclusion.-} In this work, we utilize a high-purity aluminum transmission line to connect two modules in a bosonic mode-based superconducting quantum network. \red{The single-photon relaxation time of the transmission line mode reaches 72~$\mu$s, corresponding to a quality factor of $1.7 \times 10^6$, surpassing prior low-loss interconnects while maintaining high device interchangeability.} To achieve high-quality beamsplitter interactions between the cavity and the transmission line mode, we implement SNAILs as the couplers that leverage three-wave mixing to enable efficient inter-module state transfer ($\eta=97.2\%$) and entanglement distribution (fidelity=92.5\%).
The system's compatibility with bosonic QEC is demonstrated through high-fidelity transfer of binomial logical states (fidelity = 94.8\%) between modules. Our ultra-high-$Q$ transmission line design circumvents dominant loss mechanisms in current superconducting quantum networks. 
Numerical simulations reveal the primary sources of residual infidelity and suggest mitigation strategies, advancing toward robust bosonic-mode quantum interconnects. 
\red{Our setup is compatible with longer transmission lines and, when combined with bosonic QEC codes, enables faithful quantum state transfer traversing high-temperature stages~\cite{hqst_theo,hqst_zyp,hqst_yam}. In addition, microwave-to-optical transducers can potentially be integrated to our architecture to implement quantum state transfer across remote dilution refrigerators ~\cite{zou2025}.}
	
	\smallskip{}
	
\begin{acknowledgments}
This work was funded by the Quantum Science and Technology-National Science and Technology Major Project (Grant Nos.~2024ZD0301500 and 2021ZD0300200) and the National Natural Science Foundation of China (Grants No. 92365301, 92565301, 92165209, 12061131011, 92265210, 11890704, 92365206, 12474498). This work was also supported by the Fundamental Research Funds for the Central Universities and USTC Research Funds of the Double First-Class Initiative. This work was partially carried out at the USTC Center for Micro and Nanoscale Research and Fabrication.

\end{acknowledgments}

\end{document}


\title{High-efficiency quantum state transfer for modular networks}

	\author{Hongwei Huang}
    \thanks{These two authors contributed equally to this work.}
    \author{Jie Zhou}
    \thanks{These two authors contributed equally to this work.}
    \affiliation{Center for Quantum Information, Institute for Interdisciplinary Information Sciences, Tsinghua University, Beijing 100084, China}
	\author{Weizhou~Cai}
    \affiliation{CAS Key Laboratory of Quantum Information, University of Science and Technology of China, Hefei 230026, China}
	\author{Weiting Wang}
	\affiliation{Center for Quantum Information, Institute for Interdisciplinary Information Sciences, Tsinghua University, Beijing 100084, China}
	\author{Yilong Zhou}
	\affiliation{Center for Quantum Information, Institute for Interdisciplinary Information Sciences, Tsinghua University, Beijing 100084, China}
    \author{Yunlai Zhu}
	\affiliation{Center for Quantum Information, Institute for Interdisciplinary Information Sciences, Tsinghua University, Beijing 100084, China}
    \author{Ziyue Hua}
	\affiliation{Center for Quantum Information, Institute for Interdisciplinary Information Sciences, Tsinghua University, Beijing 100084, China}
	\author{Yifang Xu}
	\affiliation{Center for Quantum Information, Institute for Interdisciplinary Information Sciences, Tsinghua University, Beijing 100084, China}
	\author{Lida Sun}
	\affiliation{Center for Quantum Information, Institute for Interdisciplinary Information Sciences, Tsinghua University, Beijing 100084, China}
 	
	  \author{Juan Song}
	\affiliation{Beijing Academy of Quantum Information Sciences, Beijing, China}
    \author{Tang Su}
	\affiliation{Beijing Academy of Quantum Information Sciences, Beijing, China}
	\affiliation{Hefei National Laboratory, Hefei, China}

    \author{Ming Li}
    \affiliation{CAS Key Laboratory of Quantum Information, University of Science and Technology of China, Hefei 230026, China}
	\author{Haifeng Yu}
    \email{hfyu@baqis.ac.cn}
	\affiliation{Beijing Academy of Quantum Information Sciences, Beijing, China}
	\affiliation{Hefei National Laboratory, Hefei, China}
	\author{Chang-Ling Zou}
	\email{clzou321@ustc.edu.cn}
	\affiliation{CAS Center For Excellence in Quantum Information and Quantum Physics,
		University of Science and Technology of China, Hefei, Anhui 230026,
		P. R. China.}
	\affiliation{Hefei National Laboratory, Hefei, China}
	
	\author{Luyan Sun}
	\email{luyansun@tsinghua.edu.cn}
	\affiliation{Center for Quantum Information, Institute for Interdisciplinary Information Sciences, Tsinghua University, Beijing 100084, China}
	\affiliation{Hefei National Laboratory, Hefei, China}
	
        \title{Supplementary Materials: ``High-performance quantum interconnect between bosonic modules beyond transmission loss constraints"}
	\pacs{} \maketitle
	
\section{Experimental device and setup}
\subsection{Modules}
\begin{figure}
        \centering
        \includegraphics{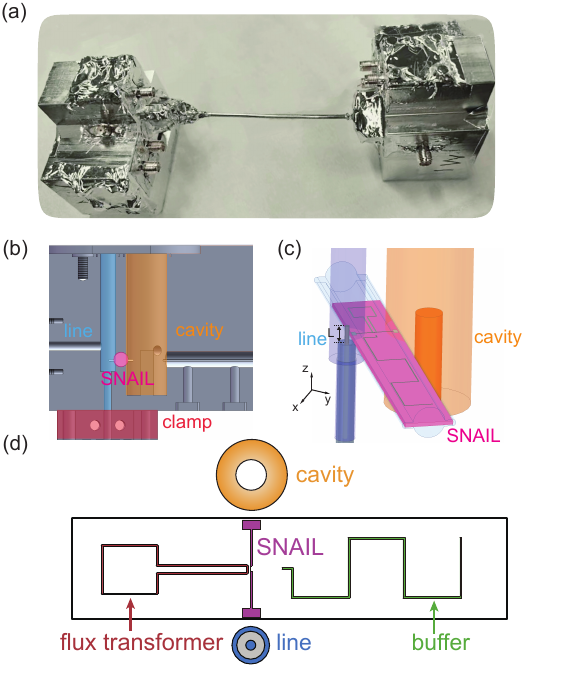}
        \caption[Schematic of the experimental device.]{\textbf{Schematic of the device.} 
            (a) Photo of the assembled modules. (b) Cross-sectional view of the Al package. The transmission line is plugged into the Al package from the bottom through a narrow tunnel with similar radius to the
            cable. (c) Schematic of the coupling between the transmission line (the bus) and the SNAIL. 
            The coupling method between the transmission line and the SNAIL is similar to that between the cavity mode and the SNAIL. (d) Top view of the sapphire chip (not to scale), including a SNAIL acting as a coupler, a flux transformer loop that delivers external flux bias generated by a coil (not shown) to the SNAIL, and a buffer mode to deliver microwave pump to the SNAIL.}		
        \label{fig:chip}
    \end{figure}

There are two nominally identical modules in our experimental setup as shown in Fig.~\ref{fig:chip}. Each module is machined from a piece of 5N5 high-purity aluminum (Al). The whole Al package is chemically etched as in Ref.~\cite{PhysRevB.94.014506} to improve the surface quality. Inside each module, there is a 3D-coaxial cavity intersected by two orthogonal trenches that house the SNAIL and transmon qubit chips. The SNAIL (pink) is coupled to a flux transformer loop for flux bias tuning and an on-chip buffer resonator (green) that delivers a strong pump, similar to the design in Ref.~\cite{PRXQuantum.4.020355}. \red{The frequency of the resonator is set close to the detuning between the cavity and the transmission line mode, acting as a band-pass filter which transmits the conversion drive while suppressing the dissipation of the high-Q cavity mode through the pump port.} The ancillary transmon qubit (not shown in Fig.~\ref{fig:chip}) is dispersively coupled to the coaxial cavity, enabling universal control of the cavity state. All chips in the experiment are fabricated in the laboratory of the Beijing Academy of Quantum Information Sciences (BAQIS) ~\cite{Wang2022TowardsPQ}. 
	
	The nonlinear coupling strengths and anharmonicities of the system are listed in Tab.~\ref{tab:SM_Hamiltonian} and the decay parameters are listed in Tab.~\ref{tab:SM_coherence}. All listed data are measured at the working bias point of the SNAILs.
    
	\begin{table}[b]
		\centering
		\caption{Measured nonlinear coupling strengths ($\chi_{ij}/2 \pi $ in MHz) and anharmonicities of the system. All values are measured at the working points.}
		\label{tab:SM_Hamiltonian}
			\begin{tabular}{c|cc} 
				\hline
				Mode   & $\text{Module 1}~(\mathrm{MHz}) $  &  $\text{Module 2}~(\mathrm{MHz})$         \\ 
				\hline
				$\text{Transmon-Cavity}$  & -1.07 & -0.94          \\
				$\text{SNAIL-Cavity}$  & -0.45  & -0.5           \\
				$\text{Transmon}$  & -222  & -234          \\
				$\text{Cavity}$  & -0.002    & -0.003        \\
				$\text{SNAIL}$  & -44  & -54            \\
				\hline
			\end{tabular}
	\end{table}
	
	\begin{table}
		\centering
		\caption{Measured mode frequencies, coherence times, and thermal populations. All values are measured at the working points.}
		\label{tab:SM_coherence}
		\begin{tabular}{c|cccc} 
			\hline
			Mode          & Frequency (GHz) & $T_1$ ($\mu$s)  & $T_2^*$ ($\mu$s) & ${n_\mathrm{th}}$ \\ 
			\hline
			Transmon 1         & 5.059           & 30                               & 40                             & 4$\%$       \\
			Cavity 1         & 5.970           & 180                               & 144                               & 3.5$\%$ 			\\
			SNAIL 1         & 4.05           & 22                                & 0.3                                & 3$\%$       \\
			Transmon 2         & 5.078           & 35                               & 58                               & 4.5$\%$      \\
			Cavity 2         & 6.092           & 148                           & 124                               & 4$\%$  \\
			SNAIL 2         & 4.137           & 26                               & 0.4                              & 4$\%$      \\
			\hline
		\end{tabular}
		\label{table:decoherence}
	\end{table}

\begin{figure}
		\centering
		\includegraphics{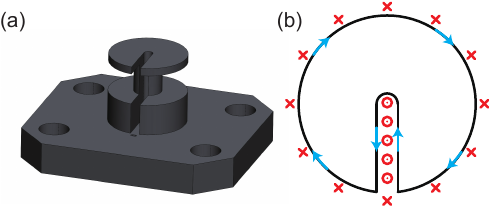}
		\caption[Schematic of the flux hose]{\textbf{Schematic of the flux hose.} (a) The flux hose is machined from a single piece of Al-6061 alloy with a small slit to guide magnetic field. The base is to secure the flux hose to the device package. Superconducting wire (not shown) is wound around the slot with a smaller diameter. (b) Bottom view diagram of the induced currents (blue) and the net magnetic field (red) generated by the induced currents and currents in the superconducting wire.
		}
		\label{fig:coil}
	\end{figure}
    
    \begin{figure*}
		\centering
		\includegraphics{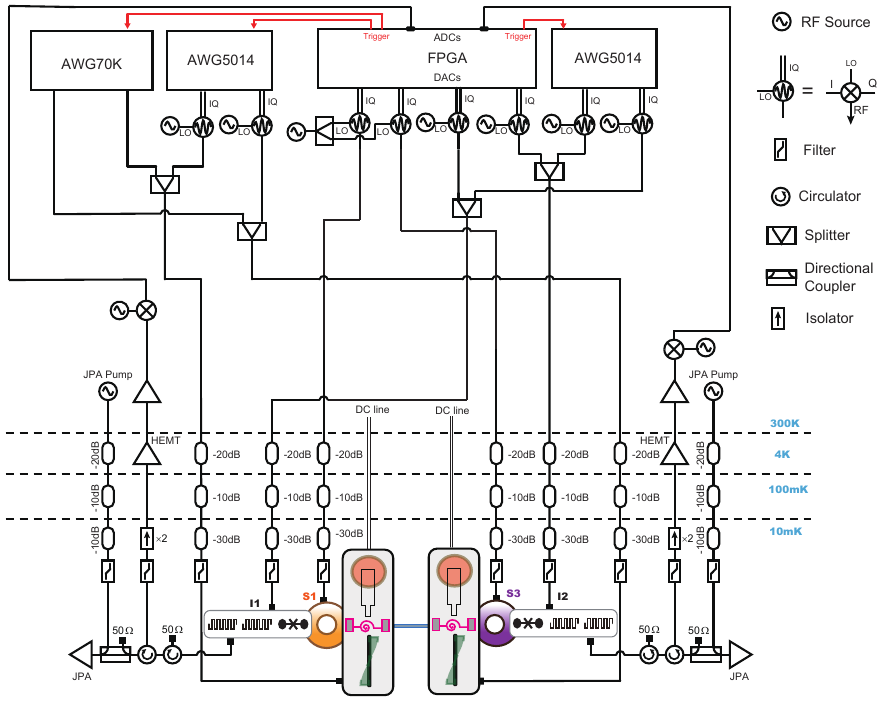}
		\caption[System wiring]{\textbf{System wiring.} 
			The input control microwave signals are generated by the single-sideband technique using IQ mixers. The LO signals are produced by microwave generators and the IQ modulation signals are generated by either AWGs or the DACs of an FPGA board. The AWGs are triggered by the FPGA to synchronize all control sequences. The LOs for the control signals for the two cavities are generated from a single microwave generator channel for phase locking. The pump signals for the two SNAILs are directly generated from two channels of an AWG70002 and the phases are naturally locked. 
		}
		\label{fig:SM_Wiring}
	\end{figure*}
	
	\subsection{Flux hose}
	Instead of a conventional coil, we employ a superconducting flux hose, machined from a single piece of Al-6061 alloy, to generate a DC magnetic field, following the design in Ref.~{\cite{flux_hose3}}. The cartoon schematic of the design is shown in Fig.~\ref{fig:coil}(a). Due to the Meissner effect, currents in the coil induce shielding currents on the surface of the flux hose, counteracting the applied magnetic field, as shown in Fig.~\ref{fig:coil}(b). These induced currents generate a magnetic field in the slit region, threading the flux transformer loop and enabling the SNAIL tuning. This design avoids the dissipation caused by the core of the coil and is easier to assemble compared to conventional flux hoses. 
	
	\subsection{Bus}

    \red{Previous experiments have shown that Al cables exhibits higher Q factors~\cite{PRXQuantum.2.030321,caoxi,niu}, so we choose to use Al cable in our experiments.} The bus mode used in our experiment is the fourth mode of an Al-PTFE-Al coaxial cable \red{from \textit{Nanjing HMC System Co.}}, with an diameter of 2.2~mm. In our setup, the length of the cable is 13~cm, with a free spectral range (FSR) of about 900~MHz. Our scheme is capable of using longer cables in the future as long as the FSR is much larger than the effective coupling strength. 
    
    At each end of the cable, the outer conductor and dielectric are carefully removed with a razor to expose inner conductor, enabling strong capacitive coupling to the SNAIL. However, a stronger coupling may lead to a lower quality factor of the bus mode. \red{In simulation, we find that longer stripped length at the end of the cable increases the energy leakage from the bus mode into the package. This, in turn, induces stronger currents through the contact area between the coaxial cable's outer conductor and the clamping fixtures at the bottom of the
    package. Given the inevitable contact resistance at the interface, the increased current introduces higher seam loss. Therefore, minimizing the stripped length serves as a practical strategy to reduce the seam loss in experimental setups. However, this introduces a trade-off: a shorter length weakens the coupling strength between the coupler and the bus mode, thereby prolonging the swap time. To overcome this limitation, we utilize the SNAIL coupler, which can realize a stronger effective beamsplitter interaction than a transmon coupler, thus compensating for the weaker capacitive coupling.} 
    
    The cable is clamped with Al brackets, similar to the design in Ref.~\cite{caoxi}. \red{In our setup, the cable is inserted from the bottom of the package through a narrow tunnel whose radius matches that of the cable, so the position in x-y plane can be precisely targeted [see Figs.~\ref{fig:chip}(b) and \ref{fig:chip}(c)]. However, the plug-in depth (the z position of the cable) may vary among each installation, leading to a coupling strength variation around 3~MHz between cooldown cycles.} 

    \red{We do not observe significant variation in the quality factor across cooldown cycles, likely because a bus with a short peeled length is less sensitive to the seam loss in the clamping region.}
	
	\subsection{Wiring}
	The detailed wiring of the whole setup is shown in Fig.~\ref{fig:SM_Wiring}. All control microwave signals are generated by the single-sideband technique with IQ mixers. For each control signal, the local oscillator (LO) signal is produced by a microwave generator and the pair of IQ signals are produced by either arbitrary waveform generators (AWGs) or digital-to-analog converters (DACs) of field programmable gate array (FPGA) cards. All AWGs are triggered by the FPGA to guarantee the synchronization. Along input lines, there are microwave attenuators and low-pass filters that are used to suppress the noise from higher-temperature plates. Readout signals are amplified on the base plate with the assistance of Josephson parametric amplifiers (JPAs). The JPAs are biased at working points with a gain of about 25~dB and a bandwidth above 10~MHz to realize high-fidelity single-shot readout of the transmon qubits. After the JPAs, there are high-electron-mobility transistor (HEMT) amplifiers at 4K plate and room-temperature amplifiers to further amplify the readout signals. 
	
    \subsection{Phase locking}
	The overall phase of the transferred states depends on the initial phase and the phases of the pumps in both modules. Therefore, the phase must be locked between the two cavity control signals and the two pump signals. This is ensured by sharing the LO for the cavity control signals in the two modules and by pumping the two SNAILs with a single AWG 70002. In contrast, the ancilla transmon and readout signals do not require phase locking and are therefore generated by individual LOs. 
	
	\subsection{Choice of mode frequencies}
	The primary consideration when choosing the frequency of each mode is $\omega_a-\omega_b=\omega_\mathrm{buf}$, where $\omega_a$, $\omega_b$, $\omega_\mathrm{buf}$ denote the cavity, bus, and buffer mode frequencies, respectively. While the buffer mode is set to the fundamental $n=1$ ($\lambda/2$) resonance, care must be taken to avoid Purcell limiting of the cavity mode by the $n=2$ ($\lambda$) mode when their frequencies are too close. 
    This consideration constrains an upper bound on $\omega_\mathrm{buf}$. On the other hand, if $\omega_\mathrm{buf}$ is too low, the influence of the $n=3$ mode ($3\lambda/2$) becomes non-negligible. As a result, we set $\omega_\mathrm{buf}/2\pi=2.4~$GHz and $\omega_a/2\pi=6.0$~GHz, with the buffer resonator's second and third mode frequencies around 4.8~GHz and 7.2~GHz, respectively, well-separated from the cavity mode. The bus mode frequency is set to $\omega_b/2\pi=3.6$~GHz, which is also far from all buffer modes. The working frequency of the SNAIL is chosen to be around 4.0~GHz to avoid hybridization with other modes while maintaining strong three-wave mixing.
    
\subsection{\red{Comparison between transmon and SNAIL}}
\red{In this work, we employ SNAILs as the coupler to control the coherent conversion between cavities and the bus based on the following two considerations.}

\red{First, compared to transmons, SNAILs offer a higher conversion rate. For a typical transmon, the four-wave mixing strength is $$g_\mathrm{BS}= E_c \frac{g_a}{\Delta_a}\frac{g_b}{\Delta_b}|\xi|^2,$$ where $E_c$ represents the Kerr non-linearity of the coupler, and $g_{a(b)}$ and $\Delta_{a(b)}$ represent the coupling strength and the detuning between the coupler and the cavity (bus) modes, respectively. $\xi=\epsilon/(\omega_p - \omega_c)$ is the reduced strength of the microwave pump. A typical value of $E_c$ for a transmon coupler is around 74~MHz~\cite{yvone_fourwave}. On the other hand, the SNAIL-induced three-wave mixing strength is $$g_\mathrm{BS}= 6g_3\frac{g_a}{\Delta_a}\frac{g_b}{\Delta_b}|\xi|.$$
For the SNAIL in our experiment, $6g_3$ can reach 300~MHz. Consequently, for given $g_{a(b)}$ and  $\Delta_{a(b)}$, the use of a SNAIL can generate a stronger $g_\mathrm{BS}$.}

\red{Second, to enhance the coupling with other modes, transmons typically exhibit large cross-Kerr interactions. When a transmon is thermally excited during drives, this cross-Kerr effect causes frequency shifts in the coupled 3D cavity and bus modes, leading to dephasing. In contrast, the SNAIL can operate at a cross-Kerr-free point, significantly mitigating such errors.}

\red{While the SNAIL offers distinct advantages, it also introduces certain limitations. For instance, the SNAIL has a short $T_\phi$, which can Purcell-limit the $T_\phi$ of the cavity and bus modes. Using alternative couplers with relatively long $T_\phi$, such as SQUIDs, might be helpful~\cite{flux_pump}. Additionally, since our current architecture lacks a dedicated readout cavity for the SNAIL, its calibration and purification become more complicated. This aspect should also be optimized in future designs.}

\section{experimental techniques}
\subsection{Readout property of the transmon qubits}
In the experiment, the readout performance will largely affect the overall quantum operation fidelity. With the assistance of JPAs, we can realize high-fidelity single-shot measurement of the transmon qubits. The readout fidelity is calibrated by first post-selecting a ground state $\left| g \right\rangle$ of the qubit, and then applying a $\pi$ pulse or no $\pi$ pulse to prepare the $\left| e \right\rangle$ or $\left| g \right\rangle$ state, followed by a second measurement. The readout fidelities $F_\mathrm{g}$ and $F_\mathrm{e}$ can be inferred by the histograms of the second measurement results in the two cases, which are shown in Tab.~\ref{tab:Transmon_readout}.
	\begin{table}[b]
		\centering
		\caption{Transmon qubit readout fidelities. Here $F_g$ ($F_e$) denotes the readout fidelity when the qubit is in $\left| g \right\rangle$ ($\left| e \right\rangle$) state.}
		\label{tab:Transmon_readout}
			\begin{tabular}{c|cc} 
				\hline
				& Module 1  & Module 2        \\ 
				\hline
				$F_\mathrm{g}$  & 99.5\% & 99.7\%         \\
				$F_\mathrm{e}$  & 97.6\% & 98.1\%           \\
				\hline
			\end{tabular}
	\end{table}
	The primary readout errors come from 1) the mismatch between the damping rate and the dispersive coupling of the readout resonator and 2) the qubit damping effect during the readout process. To mitigate the measurement errors, we adopt a calibration matrix based on Bayes' rule, which is constructed by
	$$
	F_{\text {Bayes }}=\left(\begin{array}{cc}
		F_{\mathrm{g}} & 1-F_{\mathrm{e}} \\
		1-F_{\mathrm{g}} & F_{\mathrm{e}}
	\end{array}\right).
	$$
	
	We can invert this matrix to get the corrected measurement results as 
	$$
	P_f=F_{\text {Bayes }}^{-1} \cdot P_m,
	$$
	where $P_m$ is the uncorrected measurement result and $P_f$ is the corrected final result in the main text. This calibration and correction procedure is performed on each qubit separately.
	
	\subsection{Parity measurement fidelity}
	\label{sec:parity}
	Parity measurement of the cavity state plays an important role in our experiment. It is used in the vacuum state preparation by post-selection and Wigner tomography. The parity measurement is implemented with a Ramsey-like pulse sequence $(R_{\pi/2},\pi/\chi,R_{\pm\pi/2})$, where $R_{\pm\pi/2}$ is the unconditional qubit rotation of $\pm\pi/2$ around the y axis and $\pi/\chi$ is the free evolution sandwiched between the two $\pi/2$ pulses with $\chi$ being the dispersive coupling between the qubit and the cavity. The second $R_{\pi/2}(R_{-\pi/2})$ maps the even (odd) cavity states onto the excited state of the qubit $\left| e \right\rangle$ and the odd (even) ones onto the ground state of the qubit $\left| g \right\rangle$. In our experiment, we choose to map the even cavity states onto $\left| g \right\rangle$ because both the vacuum state and the lowest-order binomial states possess even parity and mapping them onto $\left| g \right\rangle$ shows a higher parity measurement fidelity.
	
	We calibrate the parity measurement fidelity by first preparing a Fock $\ket{n}$ state in the cavity and then carry our four consecutive parity measurements. By varying $\ket{n}$, we characterize the photon-number dependence of the measurement fidelity. Lager $\ket{n}$ will degrade the qubit rotation due to the finite bandwidth of the rotation pulses. To compensate for this problem, the qubit drive frequency is shifted by $n\chi$ to set the center of the control pulse at the shifted qubit frequency in the frequency domain. The first three measurements are used to post-select a relatively ideal even or odd parity state, and the last measurement gives the fidelity value. The parity measurement fidelities for both modules are summarized in Tab.~\ref{tab:Parity}.
    
	\begin{table}[b]
		\centering
		\caption{Parity measurement fidelities.}
		\label{tab:Parity}
			\begin{tabular}{c|cc} 
				\hline
				photon number $n$& $\mathrm{Module}$ 1  & $\mathrm{Module}$ 2         \\ 
				\hline
				$1$  & 99.0\% & 99.2\%         \\
				$2$  & 97.6\% & 97.2\%           \\
				$3$  & 96.1\% & 96.0\%           \\
				\hline
			\end{tabular}
	\end{table}
    
	\subsection{Readout and calibration of the SNAILs}
\begin{figure}
		\centering
		\includegraphics{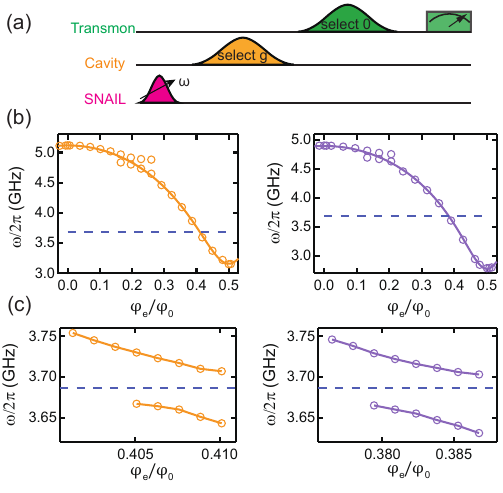}
		\caption[SNAIL calibration]{\textbf{SNAIL calibration.} 
			(a) The readout protocol of the SNAIL involves both the cavity and the transmon because there is no direct readout resonator for the SNAIL. (b) Spectrum of the SNAIL in Module 1 (left) and Module 2 (right). Circles are measured data and solid lines are fit to Eq.~\ref{equ:snail fit frequency}, from which parameters of the SNAILs are extracted. Anti-crossing at around 4.8~GHz is due to the second mode of the buffer resonator. Dashed line indicates the bus mode used in the experiment. (c) Anti-crossings between SNAIL and the bus mode for Module 1 (left) and Module 2 (right), revealing coupling strengths of 28.8~MHz and 30.6~MHZ for Module 1 and Module 2, respectively, and a bus mode frequency of 3.686~GHz.}
		\label{fig:SNAILreadout}
	\end{figure}

	In our design, there is no direct readout resonator for the SNAIL, so we utilize the dispersive interaction between the SNAIL and the cavity for the SNAIL readout. Due to the dispersive interaction, the frequency of the cavity depends on the state of the SNAIL. Therefore, the SNAIL state can be detected through a sequence of conditional operations: 1) a selective displacement of the cavity conditional on the SNAIL being in the ground state, followed by 2) a selective $\pi$ rotation applied to the qubit conditional on the $\ket{0}$ state in the cavity. The qubit remains in its ground state only when the cavity is displaced, indicating that the SNAIL is in its ground state. Figure~\ref{fig:SNAILreadout}(a) shows the pulse sequence used to readout the state of the SNAIL. In our experiment, we employ pulse durations of 4000~ns for the cavity's selective displacement and 1200~ns for the qubit's selective $\pi$ rotation. 
	
	Using this readout scheme, we obtain the SNAIL spectra in the two modules, shown in Fig.~\ref{fig:SNAILreadout}(b). While the SNAILs are designed to be nominally identical, fabrication imperfections result in parameter variations between devices. For Module 1 (Module 2), the SNAIL frequency varies from {5.116 GHz (4.906 GHz) at $\varphi_e/\varphi_0=0$ to 3.150 GHz (2.791 GHz) at $\varphi_e/\varphi_0=0.5$}, where $\varphi_e$ is the external flux and $\varphi_0$ is the flux quantum. Within this frequency range, multiple modes hybridize with the SNAILs, producing anti-crossings in the spectra. A notable example occurs near 4.8~GHz, where the buffer resonator's second mode couples to the SNAIL. We can extract the bare frequency of the SNAIL, the coupling strength, and the bare frequency of the second mode of the buffer resonator by fitting the anti-crossing data to:
	\begin{equation}
		\begin{aligned}
			\label{equ:snail bare frequency}
			& \omega_{1 0}=\omega_1-\frac{g^2}{\omega_1-\omega_0}, \\
			& \omega_{2 0}=\omega_2-\frac{g^2}{\omega_2-\omega_0},
		\end{aligned}
	\end{equation}
	where $\omega_1$ and $\omega_2$ are the measured frequencies of the two branches of the anti-crossing, $g$ is the coupling strength, $\omega_0$ is the bare frequency of the second buffer resonator mode, and $\omega_{10}$ and $\omega_{20}$ are the fitted bare frequencies of the SNAILs. 
    
    This fitting procedure can also be used to extract the coupling strength and the bare frequency of the bus mode at about 3.7~GHz, as shown in Fig.~\ref{fig:SNAILreadout}(c). Because the coupling strength of the SNAIL to the bus mode is much smaller, the anti-crossing pattern is not shown in the full spectrum. The extracted bare frequency of the bus mode is 3.68 GHz, and the coupling strengths of the SNAIL to the bus mode are 28.8 MHz and 30.6 MHz in Module 1 and Module 2, respectively.
	
	Given the bare frequency of the SNAIL, we can extract the SNAIL's other parameters by fitting the data to the following equations:
	\begin{equation}
		\begin{gathered}
			\label{equ:snail fit frequency}
			\omega_s=\sqrt{8 p c_2 E_C E_J}, \\
			c_2=\beta \cos \left(\varphi_m-\varphi_e\right)+\frac{1}{3}\cos \left(\frac{\varphi_m}{3}\right), \\
			p=\frac{c_2 E_J}{E_L+c_2 E_J}.
		\end{gathered}
	\end{equation}
	Here, $E_C$, $E_J$, and $E_L$ denote the charging energy, Josephson energy, and linear inductance energy, respectively; $\beta$ is the ratio of Josephson energies for the SNIAL's small and large junctions; $\varphi_e$ is the phase difference induced by the external flux; $\varphi_m$ is the phase difference across the SNAIL when the potential energy reaches its minimum at a given external flux, which is determined by solving 
	\begin{equation}
		\frac{d U / d \varphi}{E_J}=\beta \sin \left(\varphi-\varphi_e\right)+\sin \left(\frac{\varphi}{3}\right)=0.
	\end{equation}
	The extracted parameters for the SNAILs in the two modules are listed in Tab.~\ref{tab:snail_param}.
	\begin{table}[b]
		\centering
		\caption{Extracted parameters of SNAILs. $\beta$ is the ratio of Josephson energies for the SNIAL's small and large junctions. $\omega_b$ denotes the bare frequency of the bus's standing mode used in the experiment and $g_{sb}$ denotes the coupling strength between the SNAIL and the bus mode.}
		\label{tab:snail_param}
			\begin{tabular}{c|cc} 
				\hline
				Parameter   & Module 1  & Module 2         \\ 
				\hline
				$\beta$  & 0.162 & 0.184          \\
				$E_J/2\pi$  & 54 GHz  & 50 GHz           \\
				$E_L/2\pi$  & 165 GHz  & 149 GHz          \\
				$E_C/2\pi$  & 142 MHz    & 136 MHz        \\
				$\omega_{b}/2\pi$  & 3.686 GHz  & 3.686 GHz            \\
				$g_{sb}/2\pi$  & 28.8 MHz  & 30.6 MHz            \\
				\hline
			\end{tabular}
	\end{table}
	\subsection{Cross-Kerr between the SNAILs and the cavities}
	Although SNAIL as a coupler has the advantage of vanishing cross-Kerr term if tuned to the appropriate working point, in our design, due to imperfection in fabrication, the SNAIL at the designed working point around 4.0 GHz shows a residual cross-Kerr interaction with the cavity. The cross-Kerr value can be measured by comparing the cavity frequencies when the SNAIL is excited or not, similar to the number-splitting experiment that is used to calibrate the cross-Kerr term between the cavity and the ancillary transmon. In our experiment, the measured dispersive coupling strength between the SNAIL and the cavity is $\chi_\mathrm{sc}/2\pi=0.45$~MHz in Module 1 and $\chi_\mathrm{sc}/2\pi=0.50$~MHz in Module 2.
    
	\subsection{Reset of the SNAILs}
	The thermal excitation of the SNAIL can affect the beamsplitter performance because of the non-zero dispersive coupling. For example, if the SNAIL is initially in its excited state $\left| e \right\rangle$, the cavity's frequency would be shifted by $\chi_\mathrm{sc}$. Therefore, the conversion pump would no longer be in resonance. In addition, the frequency shift of the cavity would also lower the state encoding and decoding fidelities.
	
\begin{figure}
		\centering
		\includegraphics{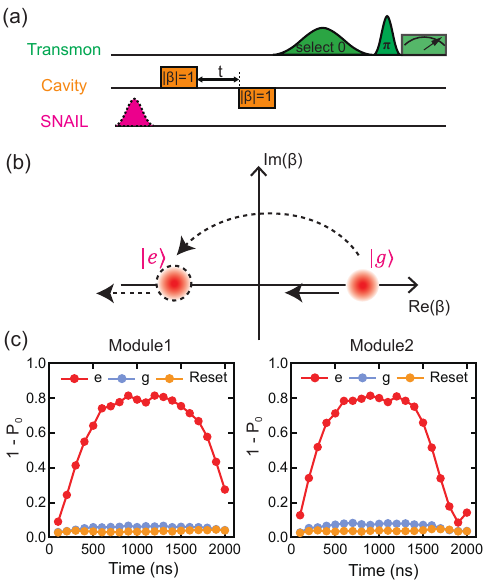}
		\caption[SNAIL purification]{\textbf{SNAIL Readout.} 
			(a) Experimental sequence used to calibrate the SNAIL readout. The principle is shown in (b). After the cavity is displaced, its state rotates around the origin if the SNAIL is in the excited state $\ket{e}$ due to the cross-Kerr interaction. After $t=\pi/\chi_\mathrm{sc}$ , the cavity state ends in $\left| -\alpha \right\rangle$ if the SNAIL is excited, otherwise it stays in $\left| \alpha \right\rangle$. With another reverse displacement, the SNAIL's population is reflected by the possibility of the cavity staying in vacuum, which is measured by a selective $\pi$ pulse on the transmon qubit. (c) Probability of finding the cavity in the vacuum state as a function of the delay time $t$, with the SNAIL excited (red) or not (blue). The slight increase of the population of the blue line at $t=\pi/\chi_\mathrm{sc}$ is due to the thermal excitation of the SNAIL. After reset (orange), most of the thermal populations are eliminated.}
		\label{fig:reset}
	\end{figure}

	To mitigate the influence of the thermal population of the SNAIL, we reset the SNAIL using the pulse sequence shown in Fig.~\ref{fig:reset}(a), similar to the method used in Ref.~\cite{PRXQuantum.2.030321}. The first pulse on the SNAIL is used to optionally prepare the $\ket{e}$ or $\ket{g}$ state for comparison. After the state preparation on SNAIL, the cavity is displaced to a coherent state with an amplitude $\alpha$ ($\alpha\sim 1$ in out experiment). Because of the dispersive coupling, if SNAIL is in the $\ket{e}$ state, the coherent state in the cavity will acquire a phase shift proportional to $\chi_\mathrm{sc} t$ during various delay times. Therefore, after $t=\pi/\chi_\mathrm{sc}$, the cavity state rotates $\pi$ around the origin in the phase space and the coherent state becomes $\ket{-\alpha}$ if the SNAIL is excited; otherwise it stays in $\ket{\alpha}$. After a reverse unconditional displacement with the same amplitude, the cavity would return to the vacuum state only when the SNAIL is in the $\left| g \right\rangle$ state. Consequently, we can post-select the SNAIL's $\ket{g}$ state by post-selecting the vacuum state of the cavity using conditional rotation of the qubit. 
    
    The diagram of the experimental principle is shown in Fig.~\ref{fig:reset}(b). Figure~\ref{fig:reset}(c) gives the probability $P_0$ of finding the vacuum state in the cavity. When the SNAIL is in $\ket{e}$ state, the cavity evolves to $\ket{-\alpha}$ for a delay time $t=\pi/\chi_\mathrm{sc}$ at which point $1-P_0$ reaches maximum. After that, the cavity will rotate back to $\ket{\alpha}$ and then return to the vacuum state after the reverse displacement, so $1- P_0$ would drop. When the SNAIL is not excited on purpose, only thermal excitation of the SNAIL would lead to rotation of the cavity state, so the curve is much flatter. After reset of the SNAIL, the curve is even flatter. In our experiment, the reset fidelity is around 0.95 for the two modules. To further suppress the effect of thermal excitation, the reset sequence can be performed twice or more times before each run of the subsequent experiment.
    
\subsection{\red{Pump-induced thermal excitation}}

\begin{figure}
		\centering
		\includegraphics{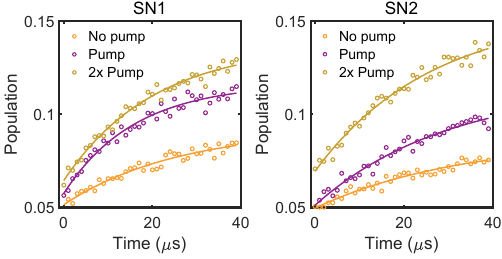}
		\caption{\red{\textbf{SNAIL's thermal excitation under microwave pumping.} Circles are measured data and solid lines are exponential fits to extract the convergence values. Due to the error inherent in this measurement method, we define the difference between the convergence value and the initial value as the thermal excitation rate of the SNAIL. An increase in the thermal excitation is observed as the pump amplitude is increased, indicating a strong dependence. The extracted thermal populations under different pump amplitudes are shown in Tab.~\ref{tab:SNAIL_thermal_table}.}}
        \label{fig:SNAIL_thermal}
\end{figure}

\red{The microwave pump used to turn on the three-wave mixing process would induce thermal excitation on the SNAIL. Because of the non-zero cross-Kerr, SNAIL's thermal excitation would shift the frequency of the coupled modes, such as cavity mode and bus mode, resulting in dephasing error. This part of error is proportional to the excitation rate of the SNAIL. Based on the method described in Fig.~\ref{fig:reset}, we measure the SNAIL's thermal excitation at different pump amplitudes and observe a positive correlation between them [see Fig.~\ref{fig:SNAIL_thermal} and Tab.~\ref{tab:SNAIL_thermal_table}]. Given that the state transfer fidelity in our experiment is not primarily limited by energy dissipation in the bus, a longer transfer time does not lead to a significant improvement. We therefore opt for a beamsplitter rate of $g_\mathrm{BS} \approx$ 0.5~MHz to balance the trade-off between the beamsplitter rate and dephasing errors.}
    
    \begin{table}[b]
		\centering
		\caption{\red{SNAIL thermal population under different pump amplitudes. The first column refers to the pump-amplitude ratio normalized to the value used in the experiment.}}
		\label{tab:SNAIL_thermal_table}
			\begin{tabular}{c|cc} 
				\hline
				Relative pump amplitude& $\mathrm{SNAIL}$ 1  & $\mathrm{SNAIL}$ 2         \\ 
				\hline
				$0$  & 4\% & 4\%         \\
				$1$  & 6\% & 6\%           \\
				$2$  & 8\% & 10\%           \\
				\hline
			\end{tabular}
	\end{table}
	\subsection{Cavity state preparation and measurement}
	All cavity states, including both Fock states and binomial states, are prepared using numerically optimized control pulses based on the GRAPE algorithm~\cite{Khaneja2005,Heeres2017}. Pulse lengths are 500~ns for the Fock-state encoding and 2000~ns for the binomial-state encoding due to the difference in the average photon numbers.
    
\begin{figure*}
		\centering
		\includegraphics{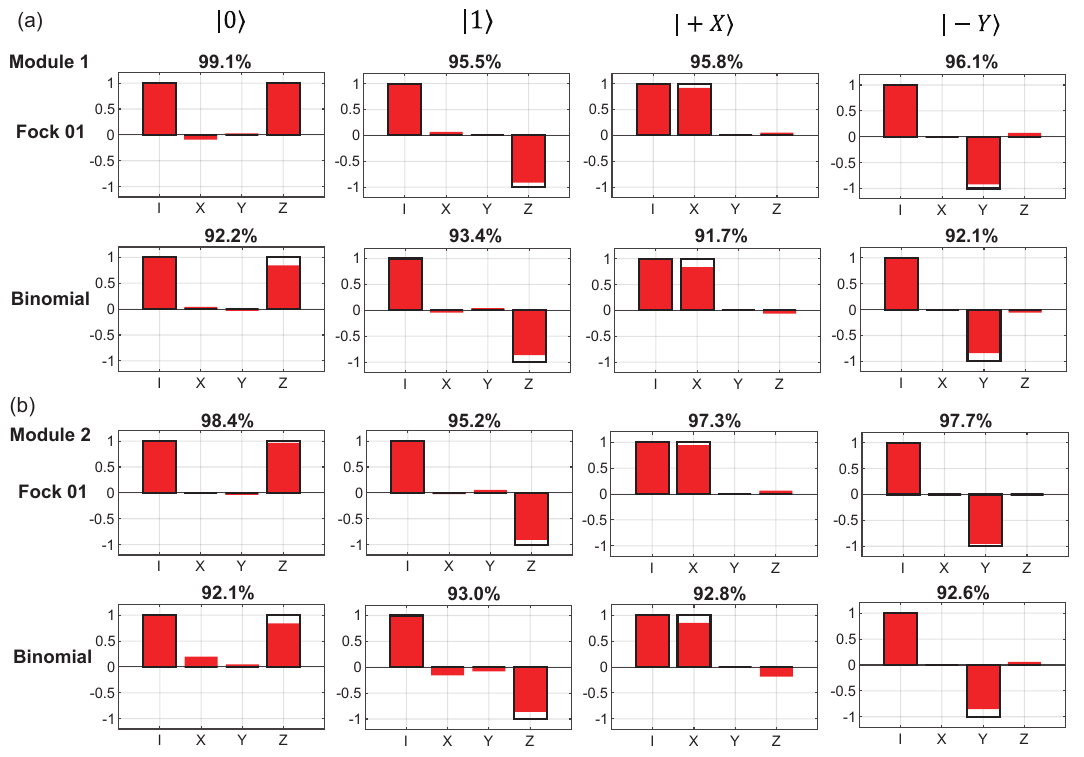}
		\caption[Initial state preparation]{\textbf{Initial state preparation and measurement.} 
			Initial state tomography for Fock $\{\ket{0},\ket{1}\}$ encoding and binomial encoding in Module 1 (a) and Module 2 (b), compared with ideal (solid) values. To perform the state tomography, we decode the quantum information onto the transmon qubit, then measure the qubit along four axes (Z, X, Y, -Z), and finally reconstruct the density matrix using MLE. These fidelities quantify the SPAM errors in the state transfer experiment.
		}
		\label{fig:ENDE}
	\end{figure*}

	There are several ways to measure the state in the cavity. To detect the photon number distribution in the cavity, one can apply conditional $\pi$ pulses to the qubit selective on the specific Fock states. However, this approach cannot reveal any phase information. A convenient way to obtain full information of the cavity state is to decode the cavity state onto the ancilla qubit using the GRAPE pulse. The results for Fock $\{\ket{0},\ket{1}\}$ and binomial encodings in the main text are based on this characterization method. The complete data for state preparations and decodings are displayed in Fig.~\ref{fig:ENDE}. The overall fidelity of this process provides a quantitative measure of the state preparation and measurement (SPAM) errors. 
	
	\subsection{Wigner tomography}
	In addition to decoding, we perform Wigner tomography to characterize the binomial states in the cavity. The measurement routine is similar to that in Ref.~\cite{Sun2014}. We first displace the cavity by a variable $\alpha$ and measure the parity using the pulse sequence described in App.~\ref{sec:parity}. To further mitigate the measurement errors due to imperfect control pulses, we perform two sequences which map the even states of the cavity to the $\ket{g}$ or $\ket{e}$ state of the ancilla qubit, respectively, and take the difference of these two resulting datasets as the parity. 
	
	Ideally, the Wigner function should theoretically integrate to unity for any physical cavity state. However, due to parity measurement errors, this condition is usually not satisfied. We normalize the measured Wigner functions by dividing the raw data by their 2D integral over the whole phase space. This normalization yields the data presented in the main text. The cavity state $\rho$ is then reconstructed from the normalized Wigner function using maximum likelihood estimation (MLE).
    
	\subsection{Tune up the conversion pump}
	In our state transfer scheme, two conditions must be satisfied: 1) the three-wave mixing strengths in the two modules must be equal, i.e., $g_1=g_2=g_\mathrm{BS}$; 2) the conversion pump must be resonant with the frequency difference between the cavity and the bus, i.e., $\omega_p=\omega_a-\omega_b$. \red{This resonant pump is optimal for this experiment because of the significantly high Q factor of the bus mode. A non-zero detuning can be employed to reduce the photon population in the bus mode during state transfer, thereby mitigating errors caused by energy dissipation in the bus~\cite{PRXQuantum.2.030321}. However, in our setup, energy dissipation in the bus contributes only approximately 0.2\% to the total state transfer error. Consequently, the primary advantage of a non-zero detuning, reducing dissipation-related errors, becomes negligible in our case. By setting $\Delta = 0$, we minimize the state transfer time without incurring a significant penalty from bus decoherence.}
    
    The conversion pump in each module can be easily tuned while keeping the beamsplitter interaction off in the other module. By fitting the patterns in Figs.~2(c) and 2(d) in the main text, we determine both the resonant pump frequency and the corresponding effective coupling strength. The resonant pump frequency typically deviates slightly from $\omega_a-\omega_b$ due to the ac-Stark shifts of the cavity and the bus, which are induced by the conversion pump. The ac-Stark shift is proportional to each mode's nonlinearity, so the shifts are different for the cavity and the bus, resulting in a slightly different resonant conversion pump frequency.

	
	The frequency of the resonant conversion pump changes when the beamsplitter interaction is turned on in the other module because the bus frequency is shifted again due to the second pump. The effect of the detuned conversion pump is complicated, but in two special cases, the syndrome is clear and obvious. In the frame of drives, the effective Hamiltonian can be written as:	
	\begin{equation}
		\frac{\hat H}{\hbar}=g_\mathrm{BS}\left( \hat a_{1} \hat b^\dagger+ \hat a^\dagger_{1} \hat b\right)+g_\mathrm{BS}\left(\hat a_{2} \hat b^\dagger+ \hat a^\dagger_{2} \hat b\right)+\Delta_{1} \hat a^\dagger_{1} \hat a_{1}+\Delta_{2} \hat a^\dagger_{2} \hat a_{2},
		\label{eq:idealH}
	\end{equation}
	where $g_\mathrm{BS}$ refers to the effective coupling strength; $\hat a_{1}$, $\hat a_{2}$, and $\hat b$ refer to the annihilation operators of cavity 1, cavity 2, and the bus, respectively; $\Delta_1$ and $\Delta_2$ refer to the pump detunings in the two modules. By solving the Heisenberg equations of motion, one can get the lossless dynamics of the system:
	\begin{equation}
		\begin{aligned}
			& \hat n_1(t)=\hat n_1(0)\left|M_{11}\right|^2+\hat n_2(0)\left|M_{12}\right|^2+\hat n_b(0)\left|M_{13}\right|^2, \\
			& \hat n_2(t)=\hat n_1(0)\left|M_{21}\right|^2+\hat n_2(0)\left|M_{22}\right|^2+\hat n_b(0)\left|M_{23}\right|^2, \\
			& \hat n_b(t)=\hat n_1(0)\left|M_{31}\right|^2+\hat n_2(0)\left|M_{32}\right|^2+\hat n_b(0)\left|M_{33}\right|^2,
		\end{aligned}
	\end{equation}
	where $\hat n_i=\hat a_i^\dagger \hat a_i$ refers to the photon population in the $i$-th mode \red{and $M_{ij}$ are the coefficients that need to be solved from Eq.~\ref{eq:idealH}.}
    
\begin{figure}
		\centering
		\includegraphics{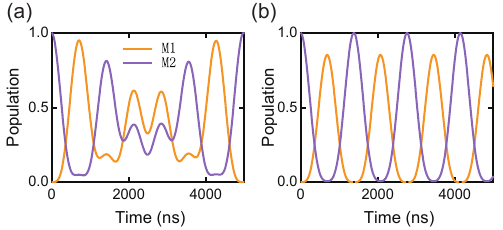}
		\caption[QuTip simulation]{\textbf{Cavity population dynamics under (a) common and (b) differential detuning of the conversion pumps.} 
			(a) Population evolution for a common detuning $\Delta_1=\Delta_2=2\pi\times 200$~kHz and $g_\mathrm{BS}=2\pi\times 492$~kHz. (b) Population evolution for a differential detuning $\Delta_1=-\Delta_2=2\pi\times 200$~kHz and $g_\mathrm{BS}=2\pi\times 492$~kHz.
		}
		\label{fig:detune}
	\end{figure}

    Suppose a single photon is prepared in Module 2 and ignore the thermal population in the bus mode, i.e, $ \left\langle \hat n_2(0)\right\rangle=1$ and $\left\langle \hat n_1(0)\right\rangle= \left\langle \hat n_b(0)\right\rangle=0$, we consider the time evolution of $ \left\langle \hat n_1(t)\right\rangle$ in two special cases:
    \begin{enumerate}
    \item for $\Delta_1=\Delta_2=\Delta_c$,
	\begin{equation}
		\begin{gathered}
			\left\langle n_1(t)\right\rangle=\frac{1}{4}\left(1-2 \cos \left(\frac{\Delta_c t}{2}\right) \cos \left(\frac{\Omega_1 t}{2}\right)+\cos ^2\left(\frac{\Omega_1 t}{2}\right)\right. \\
			\left.-\frac{2 \Delta_c}{\Omega_1} \sin \left(\frac{\Delta_c t}{2}\right) \sin \left(\frac{\Omega_1 t}{2}\right)+\frac{\Delta_c^2}{\Omega_1^2} \sin ^2\left(\frac{\Omega_1 t}{2}\right)\right),
		\end{gathered}
	\end{equation}
	with $\Omega_1=\sqrt{8 g_\mathrm{BS}^2+\Delta_{{c}}^2}$.
    \item for $\Delta_1=-\Delta_2=\Delta_d$,
	\begin{equation}
		\left\langle n_1(t)\right\rangle=\frac{4 g_\mathrm{BS}^4}{\Omega_2^4} \sin ^4 \frac{\Omega_2 t}{2}
	\end{equation}
	with $\Omega_2=\sqrt{2 g_\mathrm{BS}^2+\Delta_{{d}}^2}$. 
    \end{enumerate}
    The calculated populations of $n_1(t)$ and $n_2(t)$ are shown in Fig.~\ref{fig:detune}. The common detuning $\Delta_c$ results in a ``beat" pattern, while the differential detuning $\Delta_d$ limits the overall transfer efficiency to $4g_\mathrm{BS}^4/\Omega_2^4$. Note that arbitrary detunings $\Delta_1$ and $\Delta_2$ can be decomposed into a linear combination of $\Delta_c$ and $\Delta_d$. Therefore, by adjusting the common detuning and differential detuning we can ensure the conversion pump is in resonance. In our experiment, we first minimize the common detuning by adjusting the pump frequencies synchronously in the two modules until the beat pattern disappears. Then we tune the two pump frequencies antisymmetrically to maximize the receiver cavity photon population at $t=t_\mathrm{swap}$. 
    
	\section{Error budget analysis} 
	To understand the error budget of the state transfer process, we implement numerical simulations in QuTip~\cite{Qutip} with the effective Hamiltonian Eq.~\ref{eq:idealH} and the decoherence parameters in Tab.~\ref{table:decoherence}. The error sources can be divided into four categories: decoherence in cavities, energy loss in the transmission line, beamsplitter dephasing, and pump-induced ac-Stark shift. We can determine the contribution of each type of error to infidelity by retaining only that specific type of error in the simulation, with the corresponding results shown in Tab.~\ref{tab:error buget}. 
    
\begin{figure}
		\centering
		\includegraphics{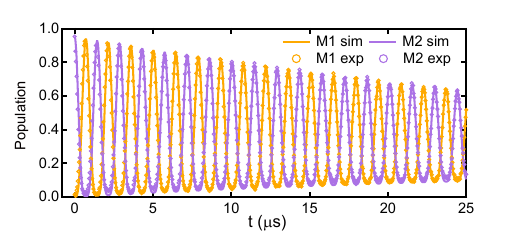}
		\caption[QuTip simulation]{\textbf{QuTip simulation results.} 
			We utilize the decoherence parameters obtained from experiments, along with the ideal beamsplitter Hamiltonian, to simulate the state transfer process in QuTiP. To emulate the dephasing induced by the pumps, we introduce a dephasing rate of $\Gamma_\Phi=1/38~\mu$s to the bus mode. The simulation results fit the experimental data well.
		}
		\label{fig:simu}
	\end{figure}

    We note that due to the long lifetime of the bus mode, the transmission line loss contributes only a small fraction of the total infidelity. In contrast, cavity decoherence and beamsplitter dephasing account for the majority of the infidelity. The beamsplitter dephasing comes from $T_{\phi}$ of the bus mode inherited from SNAILs and the pump-induced thermal excitation of SNAILs which dephases the bus and cavities because of the non-zero cross-Kerr. These two factors cause an equivalent dephasing rate $\Gamma_\Phi=1/38~\mu$s to the beamsplitting process, as shown in Fig.~\ref{fig:simu}. The remaining infidelity arises from pump-induced ac-Stark shifts, which vary over time during the pump's rising and falling edges, leading to detuning of the beamsplitter interaction from its resonance. We note that this as-Stark shift effect accumulates and becomes significant with increasing gate count as in the single-photon transfer experiment. In contrast, this effect is less pronounced in the binomial code experiment, which involves only a single round of state transfer. The residual $1\%$ infidelity for binomial code state transfer still remains unclear to us.
    
  To further improve the state transfer fidelity beyond $99\%$, it is essential to enhance the coherence time of both the cavities and SNAILs. Dephasing errors arising from pump-induced thermal excitations of SNAILs can be mitigated by biasing them at their Kerr-free points - a key advantage of using SNAILs as couplers. The time-dependent ac-Stark shift during the rising and falling edges of the pump can be addressed using the dynamic compensation techniques demonstrated in Ref.~\cite{xyf_dynamic}.
	
    \begin{table} [t]
		\centering
		\caption{Error budget of the state transfer. We quantify individual error source contributions by selectively introducing specific error types in simulations. For binomial logical states, we use the average fidelity over the three states presented in the main text as the figure of merit.}        
		\label{tab:error buget}
		\begin{tabular}{c|cc} 
			\hline
			Error type  & Fock $\{\ket{0},\ket{1}\}$ & binomial state\\ 
			& (process infidelity) &(state infidelity) \\
			\hline
			Cavity decoherence         & 0.5\%           & 1.6\%          \\
			Transmission line loss         & 0.2\%          & 0.6\%           		\\
			BS dephasing         & 0.5\%           & 1.8\%            \\
                ac-Stark shift & 0.6\%                  & 0.1\% \\
			Total         & 1.8\%          & 4.1\%                 \\
			Experiment         & 1.8\%           & 5.2\%          \\
			\hline
		\end{tabular}
	\end{table}